\documentclass[runningheads]{llncs}
\usepackage{graphicx}
\usepackage{booktabs}
\usepackage{caption}
\usepackage{subcaption}
\usepackage{array}
\newcolumntype{?}{!{\vrule width 1pt}}
\makeatletter
\newcommand{\printfnsymbol}[1]{%
  \textsuperscript{\@fnsymbol{#1}}%
}
\makeatother

\begin{document}
\title{The Pitfall of Evaluating Performance on Emerging AI Accelerators}
%
%
\author{Zihan Jiang\inst{1, 2}\thanks{Equal contribution} \and
Jiansong Li\inst{1, 2}\printfnsymbol{1} \and
Jianfeng Zhan\inst{1, 2}
}
\authorrunning{ZH. Jiang and JS. Li}
%
\institute{University of Chinese Academy of Sciences, Beijing \and
Institute of Computing Technology, Chinese Academy of Sciences, Beijing \\
\email{\{jiangzihan, lijiansong, zhanjianfeng\}@ict.ac.cn}
}
\maketitle              
\begin{abstract}
In recent years, domain-specific hardware has brought significant performance improvements in deep learning (DL). Both industry and academia only focus on throughput when evaluating these AI accelerators, which usually are custom ASICs deployed in datacenter to speed up the inference phase of DL workloads. Pursuing higher hardware throughput such as OPS (Operation Per Second) using various optimizations seems to be their main design target. However, they ignore the importance of accuracy in the DL nature. Motivated by this, this paper argue that a single throughput metric can not comprehensively reflect the real-world performance of AI accelerators. To reveal this pitfall, we evaluates several frequently-used optimizations on a typical AI accelerator and quantifies their impact on accuracy and throughout under representative DL inference workloads. Based on our experimental results, we find that some optimizations cause significant loss on accuracy in some workloads, although it can improves the throughout. Furthermore, our results show the importance of end-to-end  evaluation in DL.   

\keywords{Deep Learning \and Domain-specific Hardware \and Performance Evaluation.}
\end{abstract}
\section{Introduction}
Deep learning (DL) has revolutionized many challenge AI domains, such as image recognition \cite{he2016deep,krizhevsky2012imagenet} and natural language processing \cite{sutskever2014sequence,vaswani2017attention}. At the same time, the progressively larger DL models and datasets are getting more computationally expensive to train or inference. To keep up with the growing computational demand in modern DL workloads, hardware specialization has become a significant way \cite{hennessy2019new,jouppi2017datacenter,chen2014diannao,du2015shidiannao,liu2016cambricon}. However, maybe the gap between architecture and DL domains, these AI accelerators usually aim to provide higher throughput (e.g. OPS) and ignore the accuracy, which is the most important metric in DL. The performance numbers of mainstream AI accelerators are summarized in Table \ref{table: accelerators}. Based on our investigation, throughput is the main concern in the industry world. \emph{INT8 quantization} is their main optimization technique that brings higher throughput while saving power and memory. In addition, \emph{inference} is the most significant application of AI accelerators so far. 

\begin{table}[!ht]
\centering
\caption{The performance numbers of mainstream AI accelerators}
\begin{tabular}{|c|c|c|c|c|c|}

\toprule
\textbf{AI Accelerators} & \textbf{Producers} & \textbf{Performance Numbers} & \textbf{Memory} & \textbf{Power} & \textbf{Application} \\
\hline
TPU V1~\cite{google-tpu-v1} & Google & 92 TOPS INT8 & 8GB & 75W & Inference\\
\hline
Hanguang 800~\cite{hanguang-800} &  Alibaba & 78563 IPS INT8 & Unknown & 500 IPS/W & Inference \\
\hline
MUL100-C~\cite{cambricon-mlu100}& Cambricon & 128 TOPS INT8 & 8GB & 75W & Inference \\
\hline
Atlas 300~\cite{atlas-300}& Huawei & 64 TOPS INT8 & 32GB & 67W & Inference \\
\hline
MLU270-S4/F4~\cite{cambricon-mlu100}& Cambricon & 128 TOPS INT8 & 16GB & 70W/150W & Inference \\
\hline
TPU V2/V3~\cite{google-tpu-v2-v3}& Google & 180 TFLOPS/420 TFLOPS & 64GB/128GB & Unknown & Training and Inference \\
\toprule

\end{tabular}
\label{table: accelerators}
\end{table}
 
Previous work MLPerf \cite{mattson2019mlperf} and DawnBench \cite{coleman2017dawnbench} presents the \emph{time-to-accuarcy}, which is a metric to measure the training time to a target accuracy, to emphasis the necessity of accuracy in evaluating DL training. In this paper, we generalize the evaluation of accuracy to DL inference phase, measuring the end-to-end throughput while having accuracy constraint. From these two evaluation perspectives, we conduct a series of experiments on a typical AI accelerators, called ACC-1. Under the representative DL inference workloads, we quantifies the impact of several frequent-used optimizations on accuracy and throughput and find that some optimizations cause significant loss on accuracy in some DL workloads, although it can improves the throughout. Under the same dataset, different DL models suffer different degrees of accuracy loss. Furthermore, our results show the importance of end-to-end evaluation in DL.     

The main contribute of this paper is revealing the pitfall of evaluating emerging AI accelerators. To be specifically, a single throughput metric can not comprehensively reflect the real-world performance of AI accelerators as the accuracy is not negligible.  

\section{Background}

\subsection {Hardware Characteristics}
\begin{figure}[!ht]
\centering
\includegraphics[width=\textwidth]{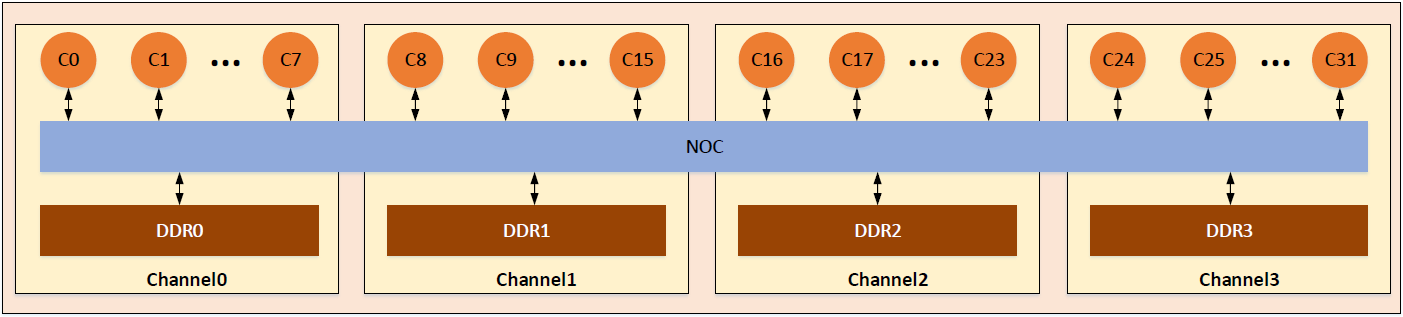}
\caption{\label{fig:cambricon-mlu100} The architectural of ACC-1}
\end{figure}

Our experiment platform is a custom ASIC deployed in datacenter to accelerate the inference phase. The general architecture is shown in Fig~\ref{fig:cambricon-mlu100}. This acclerators is based on the multi-core architecture. It includes four channels connected via a network on chip~(NOC). Each channel contains one DDR and eight computational cores. For example, Channel0 contains one DDR memory controller~(DDR0) and eight computational cores, namely C0, C1, ..., C7. DDR is responsible for the storage of DNN model, input and output of DL workloads. While those computational cores perform the execution of DNN computation tasks.
\subsection {Software Stack}  

\begin{figure}[!ht]
\centering
\includegraphics[width=0.45\textwidth]{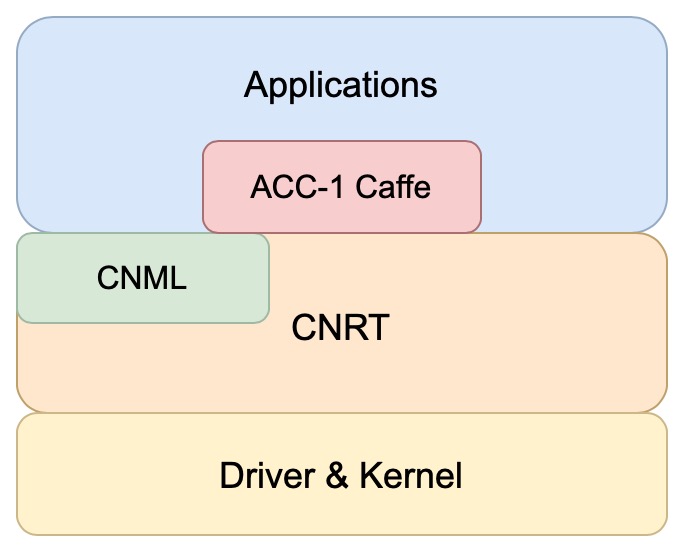}
\caption{\label{fig:cambricon-sw-stacks} Software Stacks}
\end{figure}

Fig~\ref{fig:cambricon-sw-stacks} shows the software stacks of ACC-1. As we all know, Caffe~\cite{jia2014caffe} is an open-sourced software framework used for DNN training and inference. It is written in C++ and widely adopted in research experiments and industry deployments. ACC-1 provides Caffe as its high-level programming framework. Application programmers can simply deploy their applications via ACC-1 Caffe.
CNRT is the runtime toolkit of ACC-1. It provides some common low-level utility APIs, such as device and memory management, kernel launch, task queue scheduler and etc. CNML is a wrapper of CNRT. It provides some helper functions for DNN model loading and execution and common highly-tunned DNN operators, e.g., convolution and pooling operators. Driver and kernel is responsible for the handling of memory management and interrupts.

\subsection {Frequently-used Optimizations}
The software stack provides some common utilities for optimization, such as parallelism, model parallelism and data pipeline. These approaches are not mutually exclusive.

\begin{figure}[!ht]
\centering
\includegraphics[width=0.75\textwidth]{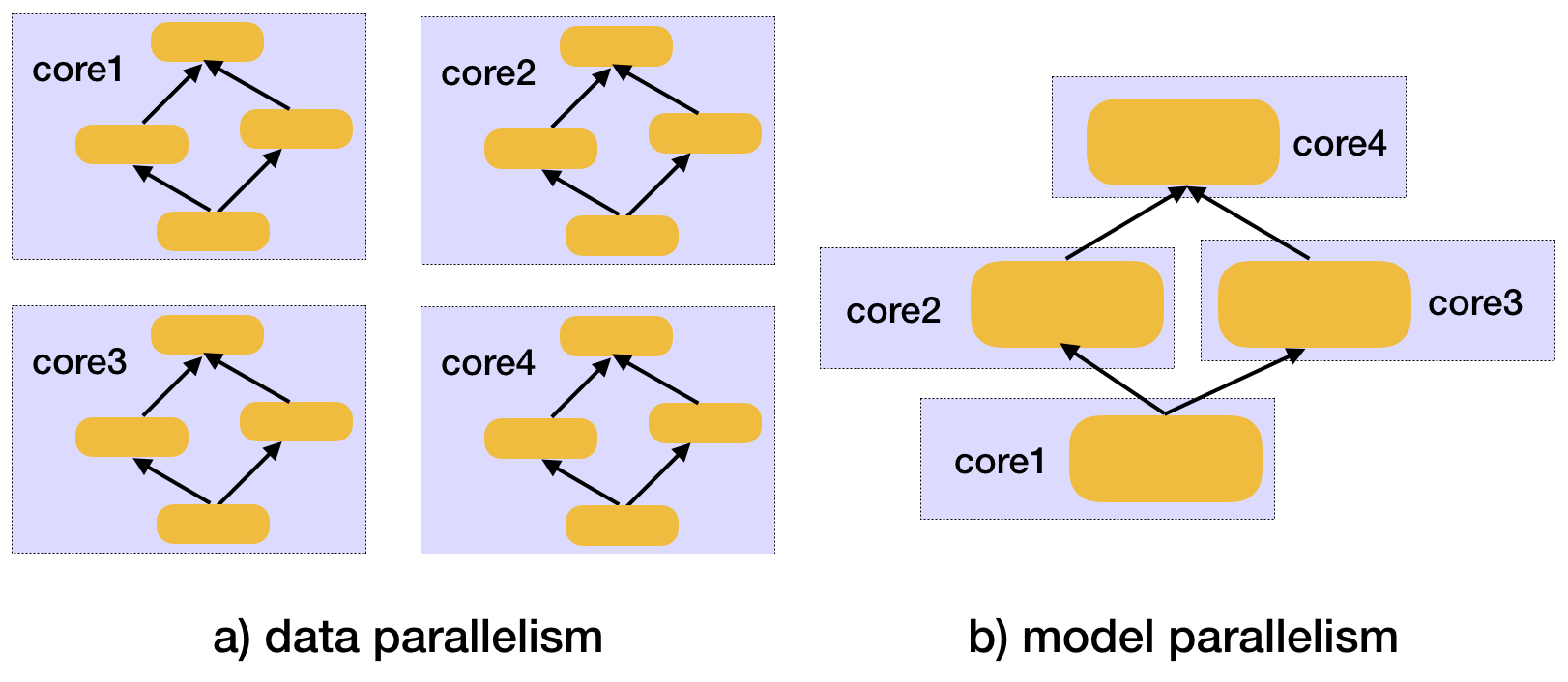}
\caption{\label{fig:dp-vs-mp} Illustration of data parallelism and model parallelism.}
\end{figure}

\noindent
{\textbf{Data Parallelism.}} In the inference phase of DL workloads, data parallelism means that given a CNN model, the input data is partitioned and assigned to different computational cores. As is shown in Fig.~\ref{fig:dp-vs-mp}(a), different cores have a complete copy of the DNN model. Each core simply gets a different part of the input data, and results from each core are somehow combined to get the final output. Data parallelism can greatly improve the throughout, since different parts of the input data can be executed concurrently.

\noindent
{\textbf{Model Parallelism.}} As is shown in Fig.~\ref{fig:dp-vs-mp}(b), model parallelism means that different cores are responsible for the computations of different parts in a single network. For example, each layer in the neural network may be assigned to a different core. In the DL domain, we can divide a neural network into several subnets, then put each subnet into different cores of ACC-1. Model parallelism can also improve the throughout, since for a single input, different parts of the DNN model can be executed concurrently.

\noindent
{\textbf{Data Pipeline.}} In the inference phase of DL workloads, the input data flow will be fetched into host memory from the disk, and then they will be transferred into the device memory of ACC-1. Finally they will be feed to the computational cores of ACC-1. In this case, data pipeline can improve the workload balance of data prefetching, transferring and data feeding.

\begin{figure}[!ht]
\centering
\includegraphics[width=0.75\textwidth]{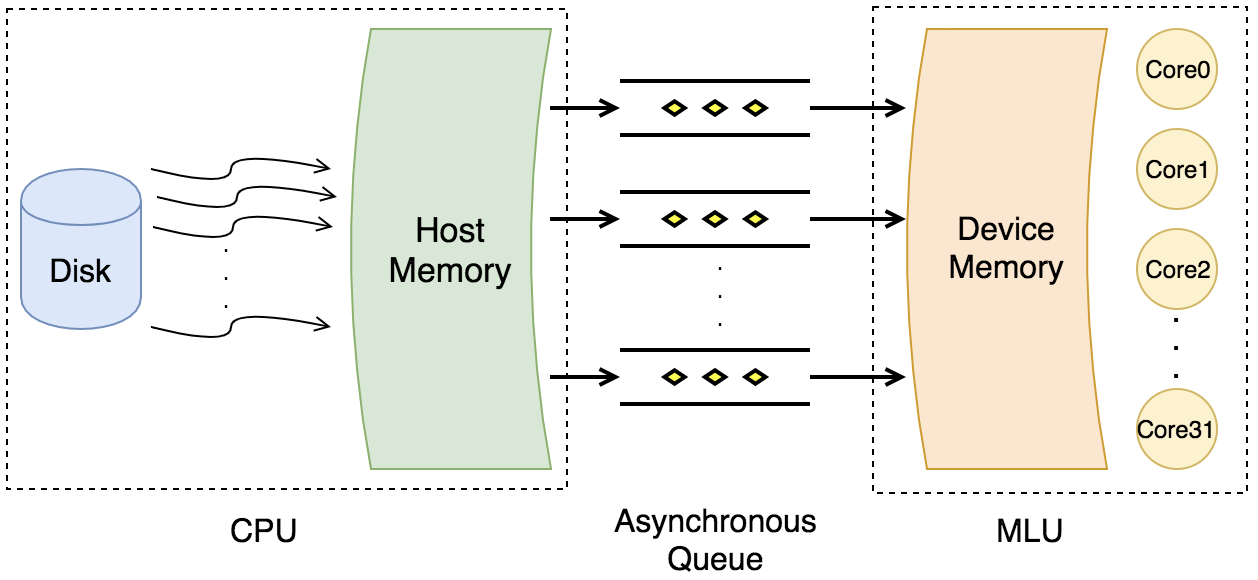}
\caption{\label{fig:data-pipeline} Illustration of data pipeline. Note that to improve the throughput, we can launch multiple threads to read data from disk into CPU memory and then dispatch the computational tasks into a queue that will be executed asynchronously. For those computational tasks within the same queue, they will be executed by their dispatching FIFO order. While those inter-queue tasks will be executed concurrently.}
\end{figure}

\noindent
{\textbf{Weights Pruning and Quantization.}} As the large amounts of synaptic weights incur intensive computation and memory accesses in the inference phase of DL workloads, researchers have proposed a number of effective techniques to explore the sparsity of DNN, including weight pruning, model compression and quantization~\cite{Zhang:2016:CAS:3195638.3195662,han2016eie,han2015deep_compression,Hubara:2017:QNN:3122009.3242044}. ACC-1 tries to exploit the sparsity and irregularity of DNN models for the performance and power efficiency. It provides tools to prune weights of input DNN model and quantize the DNN weights into low-precision fixed-point numbers, e.g., INT8. We will discuss the effects of these optimization techniques over performance and accuracy in section~\ref{sec: performance}.

\section{Performance Numbers} \label{sec: performance}

This section presents our performance numbers. As this work focus on the evaluation of accuracy in DL inference, we only analysis the optimizations that influence the accuracy. To be specifically, \emph{INT8 quantization} and \emph{weight pruning} are our main evaluation targets. Other optimizations such as parallelism and data pipeline, we don't provide specific analysis here. In the after-mentioned sections, our experiments are under the same configuration in term of other optimization techniques. Please note all of our experiments are implemented based on the customized caffe provided by the ACC-1 producers. The optimization details such as the implementation of quantization or pruning are beyond our scope.

\subsection{Metrics}
Our metrics are divided into two categories, namely throughput and accuracy. 
\subsubsection{Throughput Metrics.}
We adopt \emph{hardware FPS} and \emph{end-to-end FPS} to measure the throughput. FPS means frames per seconds. In our experiments, frame is essentially image. 
\subsubsection{TOP-1 Accuracy.}
This metric means that the model answer (the one with the highest probability) must be exactly the expected answer.

\subsection{Benchmark}
Our benchmark is summarized in the table \ref{table: benchmark}. In addition to the models and datasets, we also emphasis the expected accuracy of FP16 version. Note that the accuracy of these pre-trained models is not the state-of-the-art and doesn't represent the ability of corresponding models. We mainly use them to compare the accuracy of INT8 version. 

Our quantization experiments are based on the ImageNet dataset \cite{deng2009imagenet} and cover all of the models mentioned in the following table. The weight pruning experiments now only performed under the ResNet-50 and CIFAR10 dataset.

\begin{table}[!ht]
\centering
\caption{The benchmark specification.}
\begin{tabular}{|c|c|c?c|c|c|}

\toprule
\textbf{Datasets} & \textbf{Models} & \textbf{FP16 Accuracy} & \textbf{Datasets} & \textbf{Models} & \textbf{FP16 Accuracy} \\
\hline
ImageNet~\cite{deng2009imagenet} & AlexNet~\cite{nips-12-alexnet} &  55.816\% & ImageNet & ResNet-101 & 72.988\%\\
\hline
ImageNet &  GoogleNet~\cite{googlenet} & 68.002\% & ImageNet & ResNet-152 & 72.972\% \\
\hline
ImageNet & Inception-V3~\cite{inception-v3} & 71.570\% & ImageNet & SqueezeNet~\cite{squeezenet} & 57.076\% \\
\hline
ImageNet & MobileNet~\cite{mobilenet} & 67.140\% & ImageNet & VGG16~\cite{14-vgg} & 67.560\% \\
\hline
ImageNet & ResNet-18~\cite{15-resnet} & 64.710\% & ImageNet & VGG19 & 70.234\% \\
\hline
ImageNet & ResNet-34 & 71.100\% & CIFAR10/ImageNet & ResNet-50 & 73.024\%/84.390\% \\
\toprule

\end{tabular}
\label{table: benchmark}
\end{table}

\subsection{INT8 Quantization}
\subsubsection{Accuarcy.}
\begin{figure}[!ht]
\centering
\includegraphics[width=0.75\textwidth]{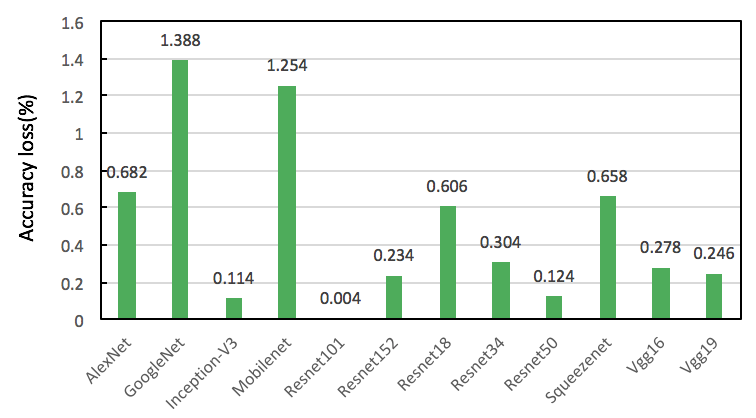}
\caption{\label{fig: accuarcy-loss} The accuracy loss after INT8 quantization compared to FP16 version.}
\end{figure}

As shown in Fig.~\ref{fig: accuarcy-loss}, the inference accuracy of all of the workloads is reduced. The loss of accuracy in AlexNet, ResNet-18 and SqueezeNet is greater than 0.6\%, which cannot be ignored in the DL field. Strikingly, the loss in GoogleNet and MobileNet achieve 1.388\% and 1.254\%. Although the results of some workloads (e.g. ResNet-101) is negligible, they are only about 60 percent of the whole benchmarking workloads. 

\subsubsection{Throughput.}

\begin{figure}[ht]
\begin{subfigure}{.5\textwidth}
  \centering
  \includegraphics[width=1\linewidth]{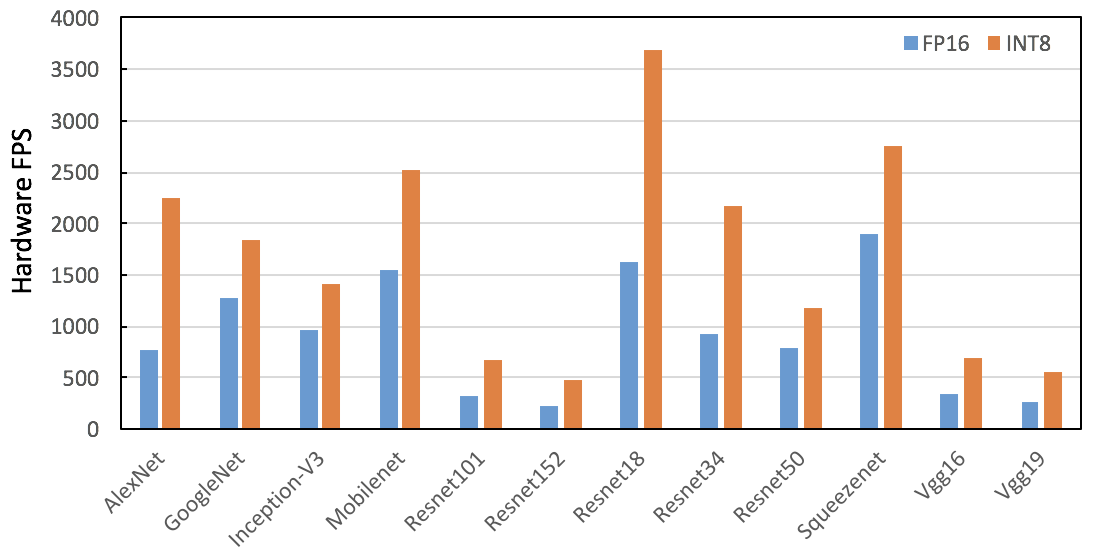}  
  \caption{The impact of INT8 quantization on hardware FPS. }
  \label{fig: int8-end-to-end-fps}
\end{subfigure}
\begin{subfigure}{.5\textwidth}
  \centering
  \includegraphics[width=1\linewidth]{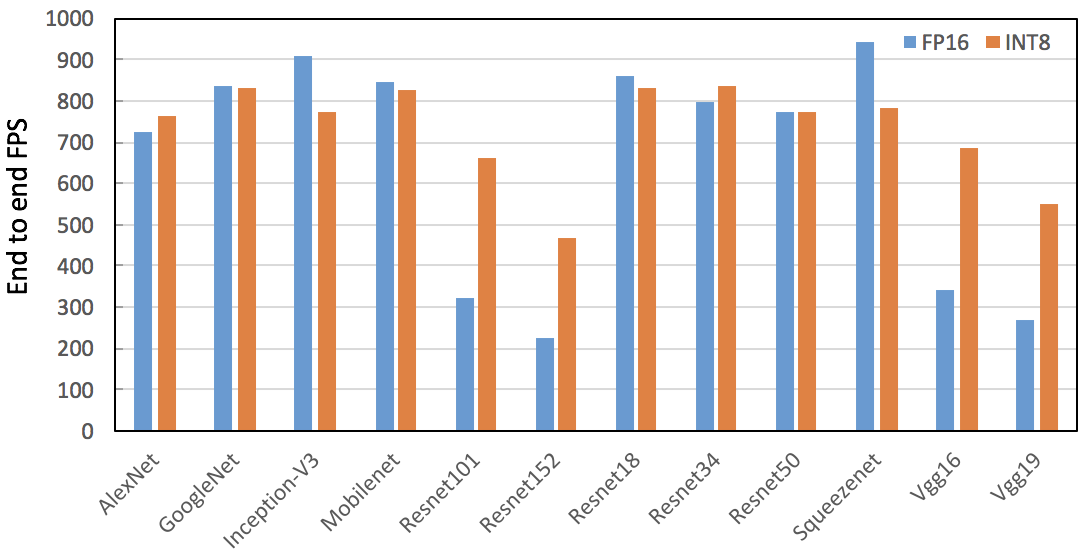}  
  \caption{The impact of INT8 quantization on end-to-end FPS.}
  \label{fig: int8-hardware-fps}
\end{subfigure}
\caption{The impact of INT8 quantization on throughput.}
\label{fig: int8-quantization}
\end{figure}

As is shown in Fig.~\ref{fig: int8-end-to-end-fps}, compared with FP16, INT8 quantization can significantly improve hardware throughput because of lower memory storage overhead. However, higher hardware throughput doesn't mean higher end to end throughput. Because the load balance of data feeding between host CPUs and AI accelerators will affect the end to end throughput directly. Therefore, we can see that in Fig.~\ref{fig: int8-hardware-fps}(b), INT8 quantization can even degrade the end to end FPS of some DNN models, e.g., Inception-V3 and SqueezeNet.

\subsection{Weights Pruning}
\begin{figure}[ht]
\begin{subfigure}{.5\textwidth}
  \centering
  \includegraphics[width=1\linewidth]{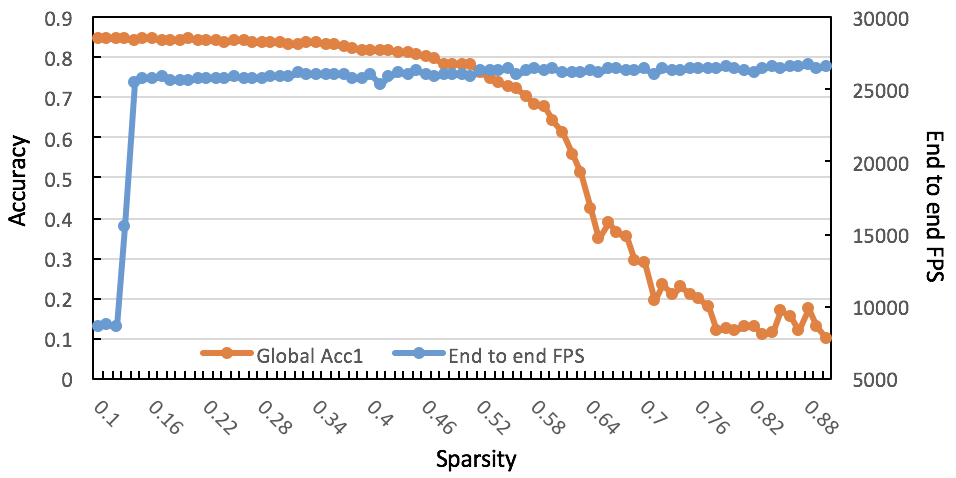}  
  \caption{The impact of sparsity on accuracy and End-to-end FPS.}
  \label{fig: pruning-end-to-end-fps}
\end{subfigure}
\begin{subfigure}{.5\textwidth}
  \centering
  \includegraphics[width=1\linewidth]{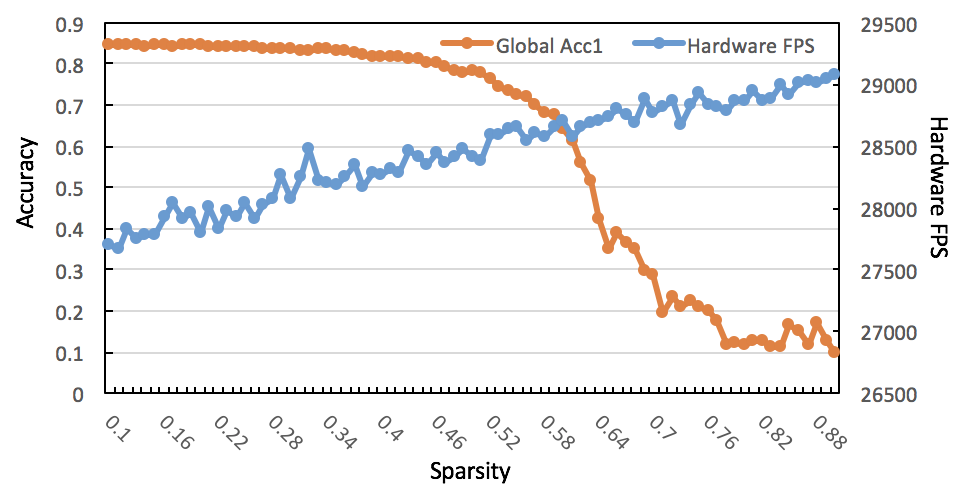}  
  \caption{The impact of sparsity on accuracy and hardware FPS.}
  \label{fig: pruning-hardware-fps}
\end{subfigure}
\caption{The impact of weight pruning.}
\label{fig: weights-pruning}
\end{figure}

As shown in Fig.~\ref{fig: weights-pruning}, weights pruning can significantly affect the end-to-end and hardware throughput. In Fig.~\ref{fig: pruning-hardware-fps}, the hardware FPS increases when the input weight sparsity is higher and higher, since higher sparsity means more zeros in the weights data and higher throughput of the accelerators. However, in Fig.~\ref{fig: pruning-end-to-end-fps}, with the weights sparsity increases, the end to end FPS improvement slows down because of the load imbalance of data feeding between host CPUs and AI accelerators.

\section{Future work}
\subsection{Workloads Variability}
Our experiments are preliminary and limited in CNN models of image recognition. In order to cover the most representative AI applications, our benchmark should include more DL workloads such as RNN and transformer \cite{vaswani2017attention}. 

\subsection{Cross-platform Evaluation}
So far, we only evaluated the optimizations on ACC-1. Cross-platform evaluation is good for finding the most suitable platform based on the models of interest. Furthermore, It's also help the producers find their design deficiencies.  

\subsection{The quality of The Pre-trained Models}
Intuitively, the quality of the pre-trained models influence our evaluation results especially accuracy to some extent. However, training the state-of-the-art models is not easy. Even if there are many state-of-the-art pre-trained models on the internet, porting them to the customized platforms of various AI accelerators is still an engineering-heavy work. 

\section{Conclusion}
We reveals a pitfall in evaluating the performance of emerging AI accelerators, that is, a single throughput metric such as OPS can not comprehensively reflect the real-world performance of AI accelerators as the accuracy is ignored. 
On a typical AI accelerator platform, we quantify the impact of INT8 quantization and weights pruning---two frequently-used optimizations for improving the throughput---on accuracy. Our results show that INT8 quantization causes significant loss on accuracy in some representative DL inference workloads. We highlight the importance of end-to-end evaluation in DL. Particularly, high hardware throughput does not mean high end-to-end throughput due to the influence of data feeding between host CPUs and AI accelerators.

%
%
%

%


\begin{thebibliography}{10}
\providecommand{\url}[1]{\texttt{#1}}
\providecommand{\urlprefix}{URL }
\providecommand{\doi}[1]{https://doi.org/#1}

\bibitem{hanguang-800}
Alibaba: Announcing {Hanguang 800: Alibaba's First AI-Inference Chip}.
  \url{https://www.alibabacloud.com/blog/announcing-hanguang-800-alibabas-first-ai-inference-chip\_595482}

\bibitem{cambricon-mlu100}
Cambricon: Cambricon {MLU}100.
  \url{http://www.cambricon.com/index.php?c=page\&id=20}

\bibitem{chen2014diannao}
Chen, T., Du, Z., Sun, N., Wang, J., Wu, C., Chen, Y., Temam, O.: Diannao: A
  small-footprint high-throughput accelerator for ubiquitous machine-learning.
  In: ACM Sigplan Notices. vol.~49, pp. 269--284. ACM (2014)

\bibitem{coleman2017dawnbench}
Coleman, C., Narayanan, D., Kang, D., Zhao, T., Zhang, J., Nardi, L., Bailis,
  P., Olukotun, K., R{\'e}, C., Zaharia, M.: Dawnbench: An end-to-end deep
  learning benchmark and competition. Training  \textbf{100}(101), ~102 (2017)

\bibitem{deng2009imagenet}
Deng, J., Dong, W., Socher, R., Li, L.J., Li, K., Fei-Fei, L.: Imagenet: A
  large-scale hierarchical image database. In: 2009 IEEE conference on computer
  vision and pattern recognition. pp. 248--255. Ieee (2009)

\bibitem{du2015shidiannao}
Du, Z., Fasthuber, R., Chen, T., Ienne, P., Li, L., Luo, T., Feng, X., Chen,
  Y., Temam, O.: Shidiannao: Shifting vision processing closer to the sensor.
  In: ACM SIGARCH Computer Architecture News. vol.~43, pp. 92--104. ACM (2015)

\bibitem{google-tpu-v1}
Google: An in-depth look at {Google’s first Tensor Processing Unit (TPU)}.
  \url{https://cloud.google.com/blog/products/gcp/an-in-depth-look-at-googles-first-tensor-processing-unit-tpu}

\bibitem{google-tpu-v2-v3}
Google: {What Makes TPU Fine Tuned to Deep Learning}.
  \url{https://cloud.google.com/blog/products/ai-machine-learning/what-makes-tpus-fine-tuned-for-deep-learning}

\bibitem{han2016eie}
Han, S., Liu, X., Mao, H., Pu, J., Pedram, A., Horowitz, M.A., Dally, W.J.:
  Eie: Efficient inference engine on compressed deep neural network.
  International Conference on Computer Architecture (ISCA)  (2016)

\bibitem{han2015deep_compression}
Han, S., Mao, H., Dally, W.J.: Deep compression: Compressing deep neural
  networks with pruning, trained quantization and huffman coding. International
  Conference on Learning Representations (ICLR)  (2016)

\bibitem{15-resnet}
He, K., Zhang, X., Ren, S., Sun, J.: Deep residual learning for image
  recognition. CoRR  \textbf{abs/1512.03385} (2015),
  \url{http://arxiv.org/abs/1512.03385}

\bibitem{he2016deep}
He, K., Zhang, X., Ren, S., Sun, J.: Deep residual learning for image
  recognition. In: Proceedings of the IEEE conference on computer vision and
  pattern recognition. pp. 770--778 (2016)

\bibitem{hennessy2019new}
Hennessy, J.L., Patterson, D.A.: A new golden age for computer architecture.
  Commun. ACM  \textbf{62}(2),  48--60 (2019)

\bibitem{mobilenet}
Howard, A.G., Zhu, M., Chen, B., Kalenichenko, D., Wang, W., Weyand, T.,
  Andreetto, M., Adam, H.: Mobilenets: Efficient convolutional neural networks
  for mobile vision applications. CoRR  \textbf{abs/1704.04861} (2017),
  \url{http://arxiv.org/abs/1704.04861}

\bibitem{atlas-300}
Huawei: {Atlas 300 AI Accelerator Card}.
  \url{https://e.huawei.com/en/products/cloud-computing-dc/atlas/atlas-300-ai}

\bibitem{Hubara:2017:QNN:3122009.3242044}
Hubara, I., Courbariaux, M., Soudry, D., El-Yaniv, R., Bengio, Y.: Quantized
  neural networks: Training neural networks with low precision weights and
  activations. J. Mach. Learn. Res.  \textbf{18}(1),  6869--6898 (Jan 2017),
  \url{http://dl.acm.org/citation.cfm?id=3122009.3242044}

\bibitem{squeezenet}
Iandola, F.N., Moskewicz, M.W., Ashraf, K., Han, S., Dally, W.J., Keutzer, K.:
  Squeezenet: Alexnet-level accuracy with 50x fewer parameters and
  {\textless}1mb model size. CoRR  \textbf{abs/1602.07360} (2016),
  \url{http://arxiv.org/abs/1602.07360}

\bibitem{jia2014caffe}
Jia, Y., Shelhamer, E., Donahue, J., Karayev, S., Long, J., Girshick, R.,
  Guadarrama, S., Darrell, T.: Caffe: Convolutional architecture for fast
  feature embedding. arXiv preprint arXiv:1408.5093  (2014)

\bibitem{jouppi2017datacenter}
Jouppi, N.P., Young, C., Patil, N., Patterson, D., Agrawal, G., Bajwa, R.,
  Bates, S., Bhatia, S., Boden, N., Borchers, A., et~al.: In-datacenter
  performance analysis of a tensor processing unit. In: 2017 ACM/IEEE 44th
  Annual International Symposium on Computer Architecture (ISCA). pp. 1--12.
  IEEE (2017)

\bibitem{krizhevsky2012imagenet}
Krizhevsky, A., Sutskever, I., Hinton, G.E.: Imagenet classification with deep
  convolutional neural networks. In: Advances in neural information processing
  systems. pp. 1097--1105 (2012)

\bibitem{nips-12-alexnet}
Krizhevsky, A., Sutskever, I., Hinton, G.E.: Imagenet classification with deep
  convolutional neural networks. In: Proceedings of the 25th International
  Conference on Neural Information Processing Systems - Volume 1. pp.
  1097--1105. NIPS'12, Curran Associates Inc., USA (2012),
  \url{http://dl.acm.org/citation.cfm?id=2999134.2999257}

\bibitem{liu2016cambricon}
Liu, S., Du, Z., Tao, J., Han, D., Luo, T., Xie, Y., Chen, Y., Chen, T.:
  Cambricon: An instruction set architecture for neural networks. In: ACM
  SIGARCH Computer Architecture News. vol.~44, pp. 393--405. IEEE Press (2016)

\bibitem{mattson2019mlperf}
Mattson, P., Cheng, C., Coleman, C., Diamos, G., Micikevicius, P., Patterson,
  D., Tang, H., Wei, G.Y., Bailis, P., Bittorf, V., et~al.: Mlperf training
  benchmark. arXiv preprint arXiv:1910.01500  (2019)

\bibitem{14-vgg}
Simonyan, K., Zisserman, A.: Very deep convolutional networks for large-scale
  image recognition (2014), \url{https://arxiv.org/abs/1409.1556}

\bibitem{sutskever2014sequence}
Sutskever, I., Vinyals, O., Le, Q.V.: Sequence to sequence learning with neural
  networks. In: Advances in neural information processing systems. pp.
  3104--3112 (2014)

\bibitem{googlenet}
Szegedy, C., Liu, W., Jia, Y., Sermanet, P., Reed, S.E., Anguelov, D., Erhan,
  D., Vanhoucke, V., Rabinovich, A.: Going deeper with convolutions. CoRR
  \textbf{abs/1409.4842} (2014), \url{http://arxiv.org/abs/1409.4842}

\bibitem{inception-v3}
Szegedy, C., Vanhoucke, V., Ioffe, S., Shlens, J., Wojna, Z.: Rethinking the
  inception architecture for computer vision. CoRR  \textbf{abs/1512.00567}
  (2015), \url{http://arxiv.org/abs/1512.00567}

\bibitem{vaswani2017attention}
Vaswani, A., Shazeer, N., Parmar, N., Uszkoreit, J., Jones, L., Gomez, A.N.,
  Kaiser, {\L}., Polosukhin, I.: Attention is all you need. In: Advances in
  neural information processing systems. pp. 5998--6008 (2017)

\bibitem{Zhang:2016:CAS:3195638.3195662}
Zhang, S., Du, Z., Zhang, L., Lan, H., Liu, S., Li, L., Guo, Q., Chen, T.,
  Chen, Y.: Cambricon-x: An accelerator for sparse neural networks. In: The
  49th Annual IEEE/ACM International Symposium on Microarchitecture. pp.
  20:1--20:12. MICRO-49, IEEE Press, Piscataway, NJ, USA (2016),
  \url{http://dl.acm.org/citation.cfm?id=3195638.3195662}

\end{thebibliography}
\end{document}